\documentclass[conference]{IEEEtran}
\IEEEoverridecommandlockouts
\usepackage{url}
\usepackage{cite}
\usepackage{amsmath,amssymb,amsfonts}
\usepackage{algorithmic}
\usepackage{graphicx}
\usepackage{textcomp}
\usepackage{xcolor}
\def\BibTeX{{\rm B\kern-.05em{\sc i\kern-.025em b}\kern-.08em
    T\kern-.1667em\lower.7ex\hbox{E}\kern-.125emX}}
\begin{document}

\title{Ray-Tracing vs. 3GPP TDL: Power Delay Profile Analysis in Outdoor-to-Indoor and Indoor Channels
\thanks{This research was supported by the U.S. Department of Commerce’s National Telecommunications and Information Administration (NTIA) under the Public Wireless Supply Chain Innovation Fund Grant Program (Award 24-60-IF2415: ASPEN - Advanced Signal Processing Enhancement for Next-Generation Open Radio Units), administered by the National Institute of Standards and Technology.}
}

\author{\IEEEauthorblockN{Julia Andrusenko}
\IEEEauthorblockA{\textit{Rampart Communications}\\
Linthicum Heights, MD \\
USA \\
jandrusenko@rampartcommunications.com}
\and
\IEEEauthorblockN{Chloe Makdad}
\IEEEauthorblockA{\textit{Rampart Communications}\\
Linthicum Heights, MD \\
USA \\
cmakdad@rampartcommunications.com}
}

% \author{
% \IEEEauthorblockN{Julia Andrusenko, Chloe Makdad}
% \IEEEauthorblockA{\textit{Rampart Communications}\\
% Linthicum Heights, MD, USA \\
% \{jandrusenko, cmakdad\}@rampartcommunications.com}
% }

\maketitle

\begin{abstract}
3rd Generation Partnership Project (3GPP) Technical Report (TR) 38.901 channel models (Releases 15-19) are widely used for physical-layer design and system-level evaluation in dense urban outdoor-to-indoor (O2I) and indoor environments. These models capture ensemble-averaged channel statistics but do not account for site-specific geometry. In this paper, we compare Power Delay Profiles (PDPs) derived from a deterministic ray-tracing model (Remcom Wireless InSite software) with those from the 3GPP TR 38.901 Tapped Delay Line (TDL) channel models. This comparative analysis is performed using a dense urban O2I scenario and a representative single-story indoor layout modeled in Washington, D.C., under matched link-distance and Non-Line-of-Sight (NLOS) conditions. All Wireless InSite PDPs are power-normalized to enable comparison of relative multipath delay structure. 

We evaluate root-mean-square (RMS) delay spread, mean excess delay, effective
maximum delay, and Kullback-Leibler (KL) distribution divergence. Results indicate that 3GPP TDL models generally exhibit longer delay spreads and often fail to capture deterministic, site-specific features such as late-arriving energy and irregular spikes. While TDL models can approximate primary channel features in some cases, their reliance on ensemble-averaged statistics rather than geometry limits their representation of fine multipath structures. We conclude that while 3GPP TDL models are suitable for large-scale system evaluation, deterministic or hybrid approaches are more appropriate for site-specific physical-layer design.
\end{abstract}


\section{Introduction}
\IEEEPARstart{T}{he} 3rd Generation Partnership Project (3GPP) Technical Report (TR) 38.901~\cite{3GPP38901Rel19} channel models are widely used for physical-layer design and system-level evaluation in dense urban outdoor-to-indoor (O2I) and indoor environments. These stochastic Tapped Delay Line (TDL) and Clustered Delay Line (CDL) models are statistically calibrated to measurement data and are designed to reproduce ensemble-average channel behavior across standardized deployment scenarios. Related geometry-based stochastic models, including WINNER II~\cite{WINNERII2008} and COST 2100~\cite{Liu2012}, similarly reproduce spatial and statistical channel behavior using stochastic cluster evolution and spatial consistency mechanisms. However, these models do not explicitly account for site-specific geometry, street topology, or precise material properties, limiting their ability to capture deterministic multipath structure in a particular physical environment.
In contrast, deterministic ray-tracing approaches model propagation directly from environmental geometry and electromagnetic material properties. Prior work has shown that deterministic ray-tracing methods can accurately reproduce measured Power Delay Profiles (PDPs), path loss, and delay spread characteristics in dense urban and indoor environments when detailed environmental geometry and material properties are available~\cite{Seidel1992},~\cite{Schaubach1992},~\cite{Zhou2017}. These differences raise questions regarding how accurately generic stochastic TDL profiles represent the complex multipath structure observed in a real urban or indoor environment.

This paper compares PDPs generated using the Remcom Wireless InSite deterministic ray-tracing platform~\cite{remcom_wi} at a carrier frequency of 3.5 GHz against the 3GPP TR 38.901 TDL-A, TDL-B, and TDL-C Non-Line-of-Sight (NLOS) channel models. The study models a 1 km × 1 km dense urban area in Washington, DC (Dupont Circle), featuring building construction materials with defined electrical properties, including concrete, asphalt, glass, and drywall. The propagation scenario includes two outdoor transmitters at 10 m height and one indoor transmitter at 1.5 m height, with a uniformly spaced 2-m receiver grid consisting of 666 first-floor receivers positioned at 1.5 m height. We did not have architectural blueprints, so a representative floorplan was replicated across all eight floors of a key building in the environment. The receiver grid was placed on the first floor of this representative building. We then selected a subset of receiver locations for which the ray-tracing simulation indicated link closure, and used the corresponding PDPs for detailed analysis.

We normalized selected Wireless InSite PDPs to the peak tap to remove link budget effects and enable comparison of the relative power distribution over delay. We then evaluated root-mean-square (RMS) delay spread, mean excess delay, effective maximum delay, and distribution divergence. For the Wireless InSite software predictions, a power threshold of $-30$~dB relative to the peak tap was applied when computing PDP statistics, including effective maximum delay. We selected a $-30$~dB threshold based on the smallest relative power level observed across the TDL-A, TDL-B, and TDL-C models. This thresholding reflects the fact that standardized 3GPP TDL models are statistical abstractions that do not preserve ultra-weak deterministic delay tails. While prior comparisons between stochastic and deterministic channel models often focus on scalar metrics such as path loss and RMS delay spread, fewer studies quantify full PDP structural differences using higher-order metrics. To address this limitation, this work utilizes Kullback–Leibler distribution divergence to explicitly quantify geometry-induced channel variance not captured by standardized stochastic channel models.

These results provide insights for physical-layer design, showing that while stochastic TDL models are suitable for ensemble-average performance evaluation, deterministic or hybrid geometry-aware models better capture site-specific multipath structure and delay-domain variability in dense urban and indoor environments. These differences are particularly relevant for system design considerations such as delay spread characterization and potential implications for inter-symbol interference (ISI) sensitivity and cyclic prefix dimensioning.

\section{Channel Modeling Framework}
\subsection{Wireless InSite Model}
To generate site-specific power delay profiles (PDPs), we used the Remcom Wireless InSite X3D shooting-and-bouncing-ray (SBR) propagation model with GPU acceleration, multithreading, and exact path calculations. The model accounts for terrain, structures, and building floorplans. Our prior studies have extensively validated this tool against measured wireless channel data. For example, in scattering-rich urban environments, predicted path loss achieved root mean square error (RMSE) values between 6.58 dB and 13.86 dB when compared against field measurements, which falls within the approximate 12--15 dB best-case RMSE range commonly reported for practical path loss models in the literature~\cite{phillips2012}.

\begin{figure}[htbp]
\centerline{\includegraphics[width=\columnwidth]{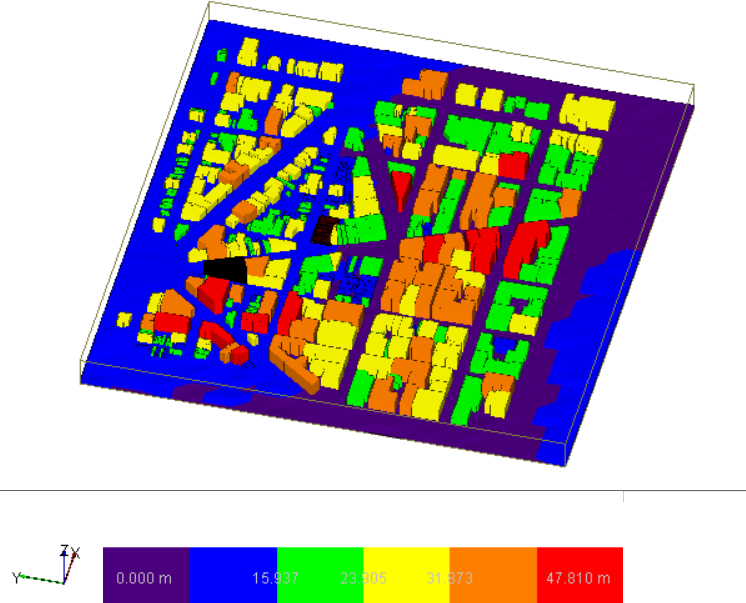}}
\caption{Wireless InSite X3D model of the Dupont Circle scenario showing terrain and building height variability through color mapping. The tallest building in the modeled environment is approximately 48~m in height.}
\label{x3d_model}
\end{figure}

The modeled scenario consisted of a 1 km $\times$ 1 km representation of the Dupont Circle area in Washington, D.C., including dense urban outdoor-to-indoor (O2I) and indoor-to-indoor (I2I) propagation conditions. Fig.~\ref{x3d_model} depicts the modeled Dupont Circle area, illustrating terrain and building height variability through color mapping, with the tallest building reaching approximately 48~m in height.

\begin{figure}[htbp]
\centerline{\includegraphics[width=\columnwidth]{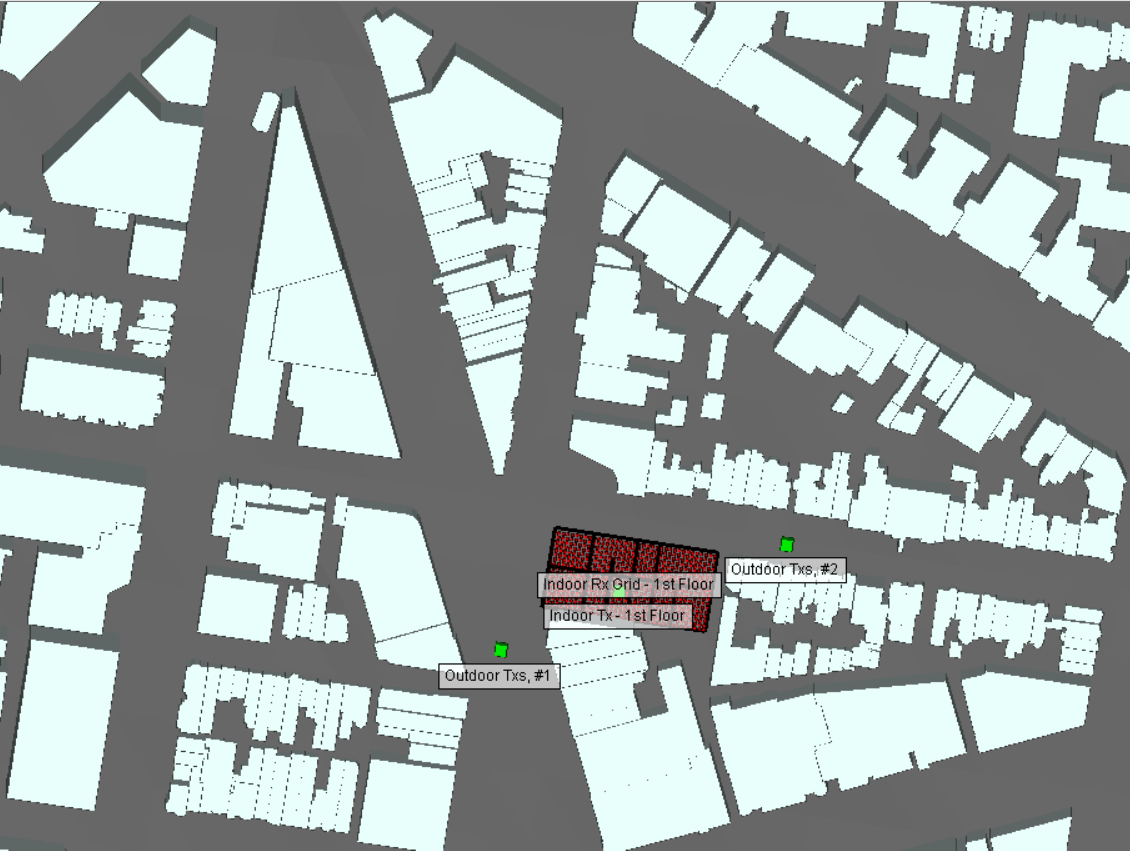}}
\caption{Wireless InSite scenario layout showing two outdoor transmitters, one indoor transmitter, and a grid of 666 indoor receivers distributed across the first floor of a representative eight-story building.}
\label{layout}
\end{figure}

Fig.~\ref{layout} depicts the Wireless InSite scenario layout, including two outdoor transmitters, one indoor transmitter, and a grid of 666 indoor receivers distributed across the first floor of a representative eight-story building. Although the building structure contains eight floors, only the first floor containing the receiver grid is shown for visualization clarity.

Wireless InSite X3D predicts multiple propagation metrics, including path loss, received power, time of arrival (TOA), time of departure (TOD), and complex impulse response. In this work, we focused specifically on the complex impulse response outputs that include absolute received powers, phases, and TOAs for each propagation path. These outputs were subsequently converted into power-normalized PDPs using excess delay rather than absolute TOA. Table~\ref{tab:wi_params} summarizes the primary modeling assumptions and simulation parameters used to generate the site-specific PDPs.

\begin{table}[t]
\centering
\caption{Wireless InSite Modeling Parameters and Assumptions}
\label{tab:wi_params}
\renewcommand{\arraystretch}{1.15}
\footnotesize
\begin{tabular}{p{0.32\columnwidth} p{0.56\columnwidth}}
\hline
\textbf{Parameter} & \textbf{Value} \\
\hline

Frequency &
3.5 GHz \\

Propagation Model &
Wireless InSite X3D SBR model \\

Environment &
1 km $\times$ 1 km Dupont Circle dense urban and indoor scenario for O2I and indoor-to-indoor (I2I) propagation modeling \\

Outdoor Transmitters (O2I Tx1 and O2I Tx2) &
10 m above ground level (AGL), below rooftop level; 0 dBi omnidirectional antennas \\

Indoor Transmitter (I2I Tx) &
1.5 m AGL with a 0 dBi omnidirectional antenna \\

Receiver (Rx) Grid &
666 indoor receivers distributed across the first floor at 1.5 m AGL with uniform 2 m spacing \\

Approximate Transmitter--Receiver Separation Distances &
O2I Tx1 to Rx grid center: 66 m \newline
O2I Tx2 to Rx grid center: 70 m \newline
I2I Tx to Rx grid center: 8 m \\

Terrain Material &
Asphalt: conductivity = $5.000\times10^{-4}$ S/m, relative permittivity = 5.72 \\

Building Materials &
Concrete: conductivity = 0.123087 S/m, relative permittivity = 5.24, thickness = 0.3 m \newline
Glass: conductivity = 0.019276 S/m, relative permittivity = 6.31, thickness = 0.003 m \newline
Two-layer drywall: conductivity = 0.027579 S/m, relative permittivity = 2.73, thickness = 0.013 m \\

Representative Building &
Eight-story building composed of concrete, glass, and drywall materials \\

Transmit Power &
Outdoor transmitters: 5 W (41 dBm) \newline
Indoor transmitter: 3.16 W (35 dBm) \newline
Representative transmit power levels were selected to ensure link closure for PDP generation. \\

\hline
\end{tabular}
\end{table}

\subsection{Processing Consistency and Thresholding}

We normalized Wireless InSite PDPs to the peak tap to remove link budget effects and enable comparison of relative delay-domain power distributions. We then applied a $-30$~dB threshold relative to the peak tap, removing weak multipath components prior to metric computations.

We computed RMS delay spread, mean excess delay, effective maximum delay, and distribution divergence on the thresholded PDPs. We selected the $-30$~dB threshold based on the minimum relative power levels observed across the TDL-A, TDL-B, and TDL-C models.

This approach reflected that standardized 3GPP TDL models are statistical abstractions that do not retain ultra-weak deterministic delay tails.

\subsection{Multipath Representation Implications}
Multipath propagation introduces critical channel impairments such as delay spread, ISI, and frequency-selective fading. The severity of these impairments directly influences system design considerations including equalization complexity and cyclic prefix sizing. Deterministic ray-traced channels can exhibit irregular delay spreads, clustered multipath behavior, and strong geometry dependence, all of which influence coherence bandwidth, ISI severity, and synchronization sensitivity. In contrast, standardized TDL models employ fixed tap structures intended to reproduce average channel behavior, which can smooth or omit fine multipath characteristics and late-arriving energy components. In addition, the TDL models are averaged over multiple channel realizations and employ fixed tap delays and relative tap powers across different delay spread regimes. However, the standard does not define a unique mapping between real environments and a specific TDL profile, so selection is based on approximate delay spread matching rather than geometry, further limiting the representation of fine multipath structure. Consequently, comparison between these modeling approaches is important, since simplified multipath representations may underestimate geometry-driven effects, particularly metrics related to tail energy and distribution divergence that are relevant to worst-case ISI assessment, cyclic prefix sizing, and site-specific physical-layer design.
\begin{table}[t]
\centering
\caption{TDL Channel Models vs. Wireless InSite X3D Model}
\label{tab:models_comparison}

\renewcommand{\arraystretch}{1.1}
\setlength{\tabcolsep}{3pt}
\footnotesize

\begin{tabular}{p{0.18\columnwidth} p{0.37\columnwidth} p{0.37\columnwidth}}
\hline
\textbf{Feature} & \textbf{TDL Models} & \textbf{Wireless InSite} \\
\hline

Modeling basis &
Statistical model from measurements &
Geometry- and physics-based propagation \\

Multipath structure &
Fixed tap delays and powers &
Explicit path-resolved components \\

Delay spread &
Ensemble-average behavior &
Geometry-dependent behavior \\

ISI characteristics &
May miss worst-case symbol overlap &
Can exhibit severe ISI from geometry-dependent multipath \\

Frequency selectivity &
Averaged fading response &
Site-specific fading response \\

Tail energy &
Limited late-arriving energy &
Preserves late multipath energy \\

Spatial dependence &
Weak geometry dependence &
Strong geometry dependence \\

Synchro\-nization sensitivity &
Smoothed timing behavior &
Geometry-driven timing variation \\

Equalization impact &
Simplified impulse response (statistical average); predictable convergence behavior &
Site-specific impulse response; geometry-dependent convergence and ISI behavior \\

Computational cost &
Low &
High \\

Representative application &
System-level evaluation &
Site-specific PHY analysis \\

\hline
\end{tabular}
\end{table}
Table~\ref{tab:models_comparison} summarizes the key differences between stochastic TDL models and deterministic Wireless InSite X3D ray-tracing model.

\section{Results}

We now compare the PDPs of two O2I urban microcell (UMi) scenarios and one I2I scenario with each of the 3GPP NLOS TDL models. For each scenario, we evaluate five different receivers. Table \ref{tab:3gpp} summarizes the delay metrics for the TDL models \cite{3GPP38901Rel19}. The RMS delay spread value (RMS) reflects the desired delay spread ($DS_{desired}$) parameter from the 3GPP standard, which is then used to compute mean excess delay (Mean) and effective maximum delay (Max), all presented in nanoseconds (ns). In Table \ref{tab:otx1}, Table \ref{tab:otx2}, and Table \ref{tab:itx}, we consider five receivers (RX ID) per scenario and display the same metrics along with KL divergence, with the site-specific models used as the reference distribution. Both the TDL and site-specific PDPs are linearized, normalized and interpolated before computing the KL divergence ($D_{KL}$) in bits.

\begin{figure}[htbp]
\centerline{\includegraphics[width=\columnwidth]{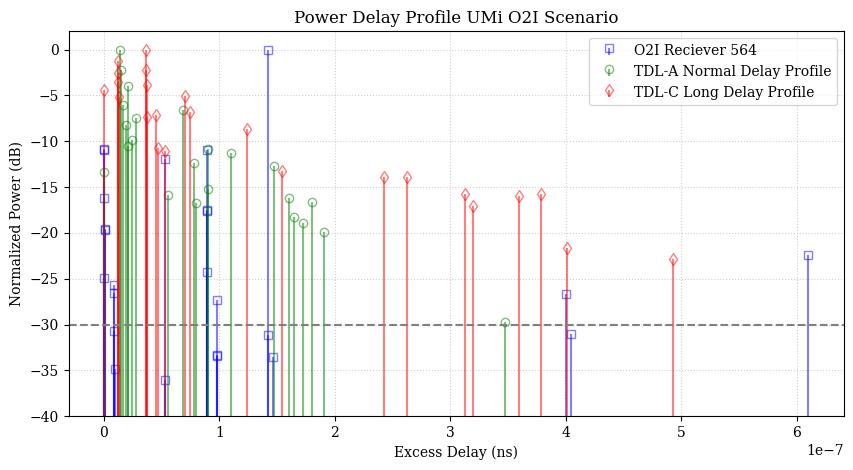}}
\caption{Power delay profiles for receiver 564 from O2I Tx1, TDL-A UMi, and TDL-C UMi.}
\label{fig:stem_oti564}
\end{figure}

\begin{figure}[htbp]
\centerline{\includegraphics[width=\columnwidth]{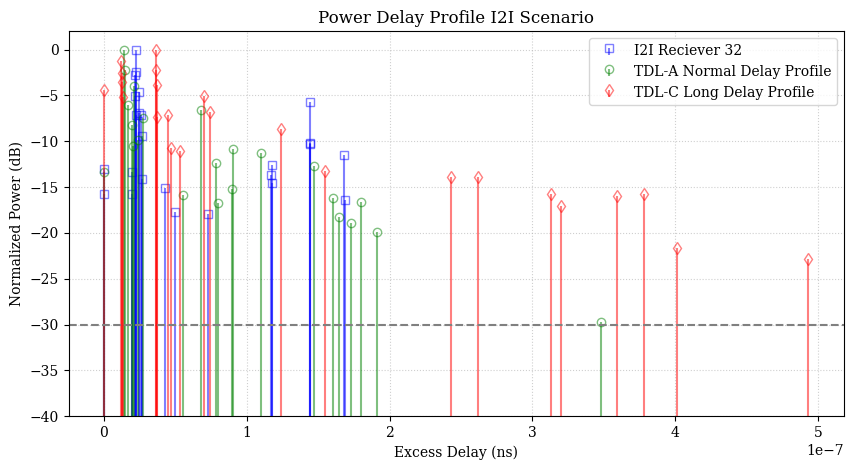}}
\caption{Power delay profiles for receiver 32 from I2I Tx, TDL-A Indoor, and TDL-C Indoor.}
\label{fig:stem_iti32}
\end{figure}

The delay spreads from the 3GPP TDL models tend to have longer tails than our site-specific models. Comparing the O2I UMi scenarios from Table \ref{tab:3gpp} to Table \ref{tab:otx1} and Table \ref{tab:otx2}, we see that the RMS delay spread, mean excess delay, and max excess delay for the 3GPP models is always longer than our site-specific ones.  When looking at the I2I scenarios in Table \ref{tab:3gpp} and Table \ref{tab:itx}, the RMS delay spread, mean excess delay, and maximum excess delay for the 3GPP models is often longer than in the site-specific ones, though this is not always the case. Figure \ref{fig:stem_oti564} and Figure \ref{fig:stem_iti32} visualize this phenomenon for O2I Tx1 receiver 564 and I2I Tx receiver 32, respectively.

\begin{table}[htbp]
\centering
\caption{3GPP TDL models}
\begin{tabular}{|c|c|c|c|c|c|}
\hline
 & \multicolumn{2}{c|}{\textbf{3GPP Model}} & \multicolumn{3}{c|}{\textbf{Delays (ns)}} \\ 
\cline{2-6} 
\textbf{Scenario} & \textbf{\textit{TDL}} & \textbf{\textit{Delay Profile}}  & \textbf{\textit{RMS}} & \textbf{\textit{Mean}} & \textbf{\textit{Max}}  \\ 
\hline
UMi O2I & A & Normal & 240 & 606.76 & 2000 \\ 
\hline
I2I & A & Normal & 36 & 91.013 & 347.71 \\
\hline
UMi O2I & B & Normal & 240 & 356.87 & 1000 \\ 
\hline
I2I & B & Normal & 36 & 53.530 &  172.20\\
\hline
UMi O2I & C & Long & 616 & 2000 & 5000 \\ 
\hline
I2I & C & Long & 57 & 147.35 & 493.18 \\
\hline
\end{tabular}
\label{tab:3gpp}
\end{table}

\begin{table}[htbp]
\centering
\caption{Wireless InSite Outdoor-to-Indoor 1}
\begin{tabular}{|c|c|c|c|c|c|c|}
\hline
 & \multicolumn{3}{c|}{\textbf{Delays (ns)}} & \multicolumn{3}{c|}{\textbf{KL Divergence (bits)}} \\ 
\cline{2-7} 
\textbf{RX ID} & \textbf{\textit{RMS}} & \textbf{\textit{Mean}} & \textbf{\textit{Max}} & \textbf{\textit{TDL-A}} & \textbf{\textit{TDL-B}} & \textbf{\textit{TDL-C}} \\ 
\hline
489 & 64.654 & 87.078 & 610.23 & 2.8673 & 3.0893 & 0.84652 \\
\hline
534 & 39.775 & 73.979 & 346.74 & 1.5593 & 0.71729 & 0.89663 \\
\hline
564 & 62.443 & 98.888 & 610.21 & 2.6485 & 2.8273 & 3.3041 \\
\hline
594 & 64.323 & 145.89 & 609.82 & 1.5414 & 0.94833 & 1.3592 \\
\hline
602 & 43.571 & 35.989 & 89.714 & 1.4708 &  0.48239 & 0.67291 \\
\hline
\end{tabular}
\label{tab:otx1}
\end{table}

\begin{table}[htbp]
\centering
\caption{Wireless InSite Outdoor-to-Indoor 2}
\begin{tabular}{|c|c|c|c|c|c|c|}
\hline
 & \multicolumn{3}{c|}{\textbf{Delays (ns)}} & \multicolumn{3}{c|}{\textbf{KL Divergence (bits)}} \\ 
\cline{2-7} 
\textbf{RX ID} & \textbf{\textit{RMS}} & \textbf{\textit{Mean}} & \textbf{\textit{Max}} & \textbf{\textit{TDL-A}} & \textbf{\textit{TDL-B}} & \textbf{\textit{TDL-C}} \\ 
\hline
484 & 56.532 & 90.464 & 335.06 & 0.58889 & 0.59527 & 0.34415 \\
\hline
521 & 81.067 & 92.487 & 334.85 & 0.57561 & 0.61502 & 0.35294 \\
\hline
559 & 80.923 & 95.142 & 334.84 & 0.55559 & 0.56111 & 0.32814 \\
\hline
595 & 45.762 & 87.650 & 335.72 & 0.64181 & 0.57692 & 0.46699 \\
\hline
631 & 44.716 & 89.606 & 335.70 & 0.83714 & 0.55084 & 0.61937 \\
\hline
\end{tabular}
\label{tab:otx2}
\end{table}

\begin{table}[htbp]
\centering
\caption{Wireless InSite Indoor-to-Indoor}
\begin{tabular}{|c|c|c|c|c|c|c|}
\hline
 & \multicolumn{3}{c|}{\textbf{Delays (ns)}} & \multicolumn{3}{c|}{\textbf{KL Divergence (bits)}} \\ 
\cline{2-7} 
\textbf{RX ID} & \textbf{\textit{RMS}} & \textbf{\textit{Mean}} & \textbf{\textit{Max}} & \textbf{\textit{TDL-A}} & \textbf{\textit{TDL-B}} & \textbf{\textit{TDL-C}} \\ 
\hline
32 & 43.930 & 62.649 & 168.84 & 1.1602 & 1.9451 & 0.61730 \\
\hline
36 & 12.796 & 39.642 & 148.07 & 2.7942 & 2.9244 & 2.0677 \\
\hline
69 & 18.262 & 32.789 & 97.297 & 0.80462 & 1.0688 & 0.59745 \\
\hline
176 & 28.131 & 126.85 & 164.45 & 1.7332 & 1.0914 & 1.6052 \\
\hline
259 & 23.507 & 39.642 & 104.02 & 2.7493 & 0.89891 & 1.7138 \\
\hline
\end{tabular}
\label{tab:itx}
\end{table}

For an approximating probability distribution $Q$ and true probability distribution $P$, KL divergence

\begin{equation*}
    D_{KL}(P \parallel Q) = \sum_{x \in \mathcal{X}} P(x) \log_2 \left( \frac{P(x)}{Q(x)} \right)
\end{equation*}
is a measure of how much $Q$ differs from $P$ \cite{kullback1951information}. KL divergence can be used to model how well an approximate distribution fits a true distribution and has previously been used to evaluate channel models versus specific communication environments \cite{kulhandjian2014modeling}. In this context, we evaluate how well the 3GPP TDL models approximate our site-specific PDPs. A KL divergence of 0 bits would mean that the 3GPP model is a perfect representation of the site-specific model, and higher KL divergence corresponds to more significant differences. Since we have computed KL divergence in terms of bits, we can interpret a $D_{KL} = 1$ as meaning that we would need one bit more per sample to represent the site-specific model with the 3GPP model than if we had simply used the site-specific model. Computing KL divergence requires us to normalize the power delay profiles, so the metric does not reflect differences in delay spread.

\begin{figure}[htbp]
\centerline{\includegraphics[width=\columnwidth]{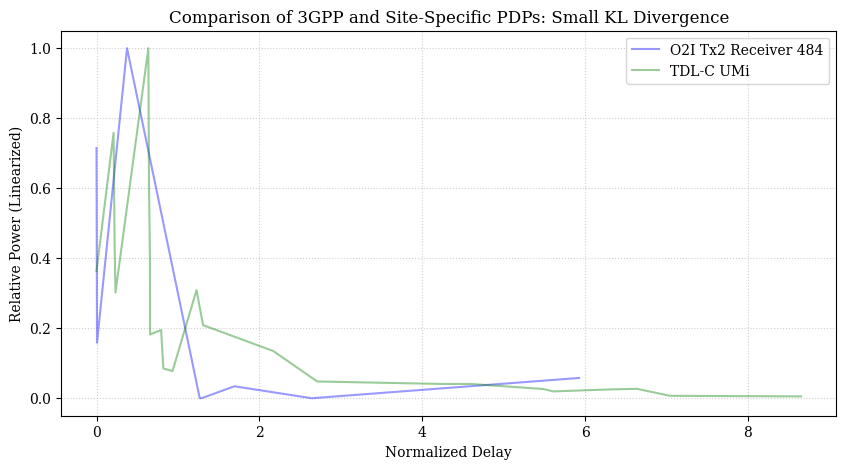}}
\caption{A comparison of the normalized PDPs of 3GPP TDL-C UMi and receiver 484 from the O2I Tx2. The KL Divergence of these PDPs is small, and the 3GPP model reflects the main features of the site-specific model.}
\label{fig:norm_pdp484}
\end{figure}

\begin{figure}[htbp]
\centerline{\includegraphics[width=\columnwidth]{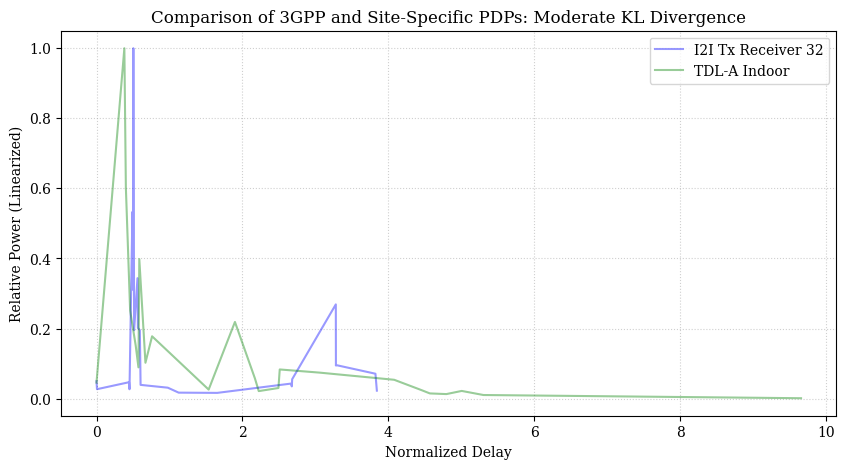}}
\caption{A comparison of the normalized PDPs of 3GPP TDL-C UMi and receiver 32 from the I2I Tx site-specific model. The KL Divergence of these PDPs is moderate, and the 3GPP model fails to capture all of the features of the site-specific model.}
\label{fig:norm_pdp32}
\end{figure}

\begin{figure}[htbp]
\centerline{\includegraphics[width=\columnwidth]{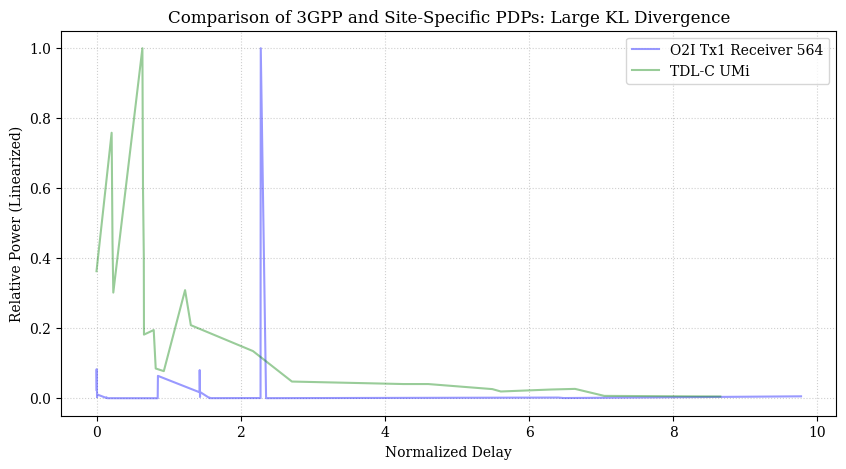}}
\caption{A comparison of the normalized PDPs of 3GPP TDL-C UMi and receiver 564 from the O2I Tx1 site-specific model. The KL Divergence of these PDPs is large, and the 3GPP model does not reflect the main features of the site-specific model.}
\label{fig:norm_pdp564}
\end{figure}

The data show many scenarios with notable differences between the 3GPP TDL models and our site-specific PDPs, particularly in our first O2I scenario and our I2I scenario. Receiver 564 in Table \ref{tab:otx1} and receiver 36 in Table \ref{tab:itx} each show massive differences with all three TDL models. Figure \ref{fig:norm_pdp564} shows the normalized PDP for receiver 564 versus the analogous PDP for the TDL-C model, and the difference between the PDPs is stark. Other receivers in both scenarios have substantial differences with some or all of the models. 

Figure \ref{fig:norm_pdp32} illustrates and example of where a 3GPP model captures some, but not all, of the features of a site-specific model. Receiver 32 from the I2I site-specific model has a spike not reflected in TDL-A, resulting in a KL  divergence value $D_{KL} = 1.16$.

There are also instances where the 3GPP models are reasonable estimates of the normalized site-specific models. The second O2I scenario, seen in Table \ref{tab:otx2}, is, in many cases, quite similar to the 3GPP models. The similarities between the PDPs TDL-C and receiver 484 are seen in Figure \ref{fig:norm_pdp484}, illustrating a scenario where 3GPP models are still capturing the primary features of the site-specific scenario.

\section{Conclusion}
This study compared the PDP characteristics of 3GPP TR 38.901 TDL models with the Wireless InSite deterministic ray-tracing simulations in dense urban O2I and I2I environments. Results demonstrate that while 3GPP TDL models exhibit longer overall delay spreads and statistical tails, site-specific models better preserve late-arriving multipath energy and deterministic structures, such as irregular power spikes, that standardized models often smooth or omit. KL divergence analysis reveals significant structural discrepancies, particularly in cases where 3GPP abstractions fail to capture primary geometry-driven features. We conclude that while 3GPP TDL models are suitable for large-scale system evaluation, deterministic or hybrid geometry-aware modeling is essential for site-specific physical-layer design, cyclic prefix dimensioning, and assessing inter-symbol interference sensitivity.

\bibliographystyle{IEEEtran}
\bibliography{bibliography}

\end{document}